\documentclass[sigconf,screen,nonacm]{acmart}
\startPage{1}
\setcopyright{none}
\acmConference[]{}{}{}
\acmISBN{}
\acmDOI{}

\usepackage[all]{nowidow}
\usepackage{xcolor}
\usepackage{subcaption}
\usepackage{hyperref}

\usepackage[capitalise]{cleveref}
\crefformat{section}{\S#2#1#3} 
\crefformat{subsection}{\S#2#1#3}
\crefformat{subsubsection}{\S#2#1#3}

\begin{document}

\author{Tobias Pfandzelter}
\affiliation{%
    \institution{TU Berlin \& ECDF}
    \department{Mobile Cloud Computing Research Group}
    \city{Berlin}
    \country{Germany}}
\email{tp@mcc.tu-berlin.de}
\author{David Bermbach}
\affiliation{%
    \institution{TU Berlin \& ECDF}
    \department{Mobile Cloud Computing Research Group}
    \city{Berlin}
    \country{Germany}}
\email{db@mcc.tu-berlin.de}

\title{\textsc{Celestial}: Virtual Software System Testbeds for the LEO Edge}

\begin{abstract}
    As private space companies such as SpaceX and Telesat are building large LEO satellite constellations to provide global broadband Internet access, researchers have proposed to embed compute services within satellite constellations to provide computing services on the \emph{LEO edge}.
    While the LEO edge is merely theoretical at the moment, providers are expected to rapidly develop their satellite technologies to keep the upper hand in the new space race.

    In this paper, we answer the question of how researchers can explore the possibilities of LEO edge computing and evaluate arbitrary software systems in an accurate runtime environment and with cost-efficient scalability.
    To that end, we present \textsc{Celestial}, a virtual testbed for the LEO edge based on microVMs.
    \textsc{Celestial} can efficiently emulate individual satellites and their movement as well as ground station servers with realistic network conditions and in an application-agnostic manner, which we show empirically.
    Additionally, we explore opportunities and implications of deploying a real-time remote sensing application on LEO edge infrastructure in a case study on \textsc{Celestial}.
\end{abstract}

\maketitle

\section{Introduction}
\label{sec:introduction}

Private aerospace and Internet companies, such as SpaceX\footnote{\url{https://www.starlink.com/}}, OneWeb\footnote{\url{https://www.oneweb.world/}}, and Telesat\footnote{\url{https://www.telesat.com/}}, are launching tens of thousands of satellites to provide global broadband Internet access.
Thanks to their low-Earth orbit (LEO) and free-space laser links, consumers can expect low-latency, high-bandwidth Internet access anywhere on Earth.
New LEO satellite networks challenge not only the old satellite-based Internet access but terrestrial fiber as well~\cite{Pultarova2015-ml,Bhattacherjee2018-vc,Sheetz_undated-ea}.

Fueled by that development, it has also been proposed to implement computing resources within these LEO constellations, e.g.,~\cite{Zhang2019-ew,Bhattacherjee2020-kr,Pfandzelter2021-dp}, to facilitate LEO edge computing, as is common with terrestrial multi-access edge computing (MEC)~\cite{etsi2022v311}.
Edge resources located on communication satellites, which act as radio uplinks for LEO Internet subscribers, could provide low-latency application access from which especially rural areas and clients without a nearby cloud data center would benefit~\cite{Bhattacherjee2020-kr,Bhosale2020-aa,pfandzelter2020edge,Pfandzelter2021-dp}.

LEO edge computing introduces novel challenges in application management:
The number of potential satellite servers, with, e.g., the proposed Starlink constellation comprising more than 40,000 satellites at completion~\cite{Handley2018-ay,Handley2019-ce,Kassing2020-yc}, will require new approaches to server management.
Further, LEO satellites move at speeds in excess of 27,000km/h and ground equipment frequently needs to reconnect to new satellites, resulting in an ever-changing network topology.
Finally, LEO edge software is subject to constrained compute resources and the harsh environment of space.

To solve these challenges, researchers will need to develop new middleware systems for application state management, request routing, and service offloading.
As LEO edge computing currently exists only as a concept and any actual infrastructure is years away from implementation, emulated testbeds that go beyond the capabilities of simulation are needed to test and benchmark such systems.
Emulating LEO edge infrastructure in the cloud, however, is non-trivial, as the number of satellite servers raises scalability and cost concerns, and the highly dynamic satellite network and environmental effects must be accurately reflected in the testbed.

In this paper, we thus raise the question of how we can evaluate \emph{arbitrary software systems} for the LEO edge \emph{as accurately as possible} and \emph{in a cost-efficient manner}.
To answer this question, we make the following contributions:

\begin{itemize}
    \item We present LEO edge computing and the challenges of building and evaluating LEO edge software without access to actual infrastructure (\cref{sec:background}).
    \item We introduce \textsc{Celestial}, a novel LEO edge emulation tool based on microVMs and discuss how it addresses these challenges (\cref{sec:systemdesign}).
    \item We evaluate these claims by deploying a latency-sen\-sitive edge application on \textsc{Celestial} (\cref{sec:videoconferencing}).
    \item In a case study on \textsc{Celestial}, we evaluate different deployments of a distributed real-time data analysis service in the context of remote sensor networks to assess the opportunities and implications of the LEO edge environment (\cref{sec:buoys}).
    \item We discuss threats to validity for our work and derive avenues for future work on \textsc{Celestial} and the LEO edge (\cref{sec:discussion}).
\end{itemize}

We make our implementation of \textsc{Celestial} available as open-source\footnote{\url{https://github.com/OpenFogStack/celestial}} to help future researchers validate their own applications and platforms.
Our hope is that this will make the field of LEO edge computing more accessible and provide a starting point for systems research in this area.

\section{Background \& Related Work}
\label{sec:background}

In this section, we give an overview of the state of the art in large LEO satellite communication networks and describe the opportunities and challenges of the novel LEO edge computing paradigm.
Furthermore, we discuss what it takes to build and test LEO edge software systems without access to LEO edge infrastructure, and where current simulators and testbeds fall short.

\subsection{Large LEO Satellite Communication Networks}

Satellite Internet access using geostationary orbits at altitudes in excess of 35,000km have been in operation for decades.
Yet, their high communication delays and low bandwidth make them infeasible for most applications~\cite{Clarke1945-qb}.
Advances in radio and laser technology~\cite{7553489}, and progressively lower satellite launch costs in the last few years~\cite{Jones2018-ct}, have now enabled private companies such as SpaceX, Telesat, OneWeb, and Amazon to deploy high-bandwidth, low-latency satellite communication networks using thousands of satellites at altitudes between 500 and 1,500km, called the low-Earth orbit (LEO).
Rather than only relaying radio signals from ground stations, these new satellites can use inter-satellite laser links (ISL) for communication between adjacent satellites.
Given the vacuum in space, these ISLs can benefit from a $\sim$47\% faster light propagation than in fiber cables~\cite{Handley2018-ay,Bhattacherjee2019-jz}.
Point-to-point communication between two ground stations over the satellite network can thus incur less communication delay than with terrestrial fiber connections.
As installing user equipment is also cheaper than installing fiber optic cables, the new large LEO satellite communication networks are expected to challenge not only traditional satellite Internet access but also terrestrial fiber~\cite{Bhattacherjee2018-vc,Bhattacherjee2019-jz}.

The core of a LEO satellite communication network is the actual satellite constellation.
Such a constellation comprises shells of satellites, each shell at a different altitude and with different orbital parameters.
Each of these shells consists of a number of orbital planes, evenly spaced around the equator.
Within each plane are satellites that follow the same orbit, evenly spaced around that plane~\cite{408677,pfandzelter_optimal:_2022}.

\begin{figure}
    \centering
    \includegraphics[width=0.95\columnwidth]{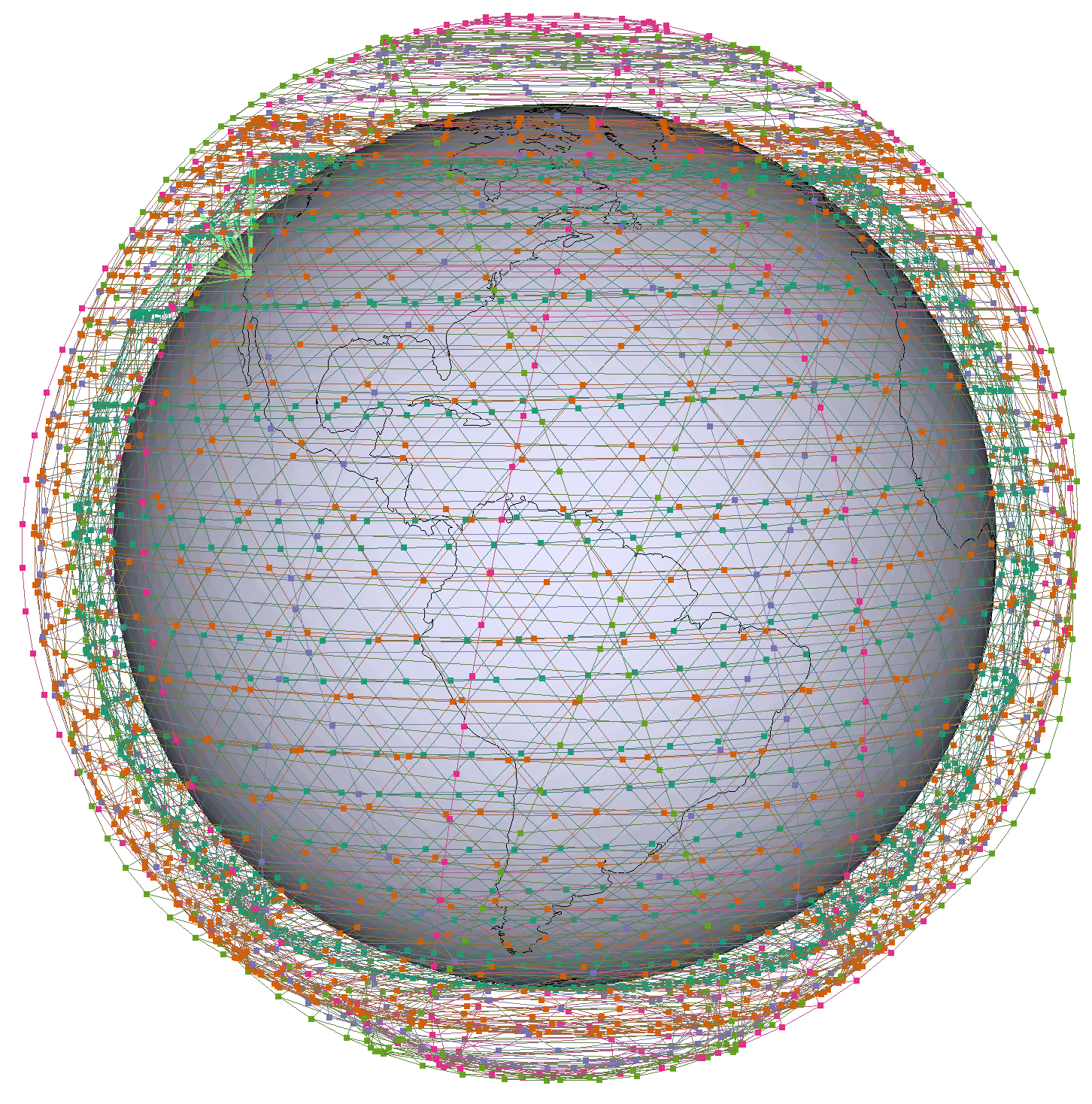}
    \caption{Overview of the planned phase \textrm{I} Starlink constellation with five shells of 1,584 satellites at 550km (turquoise), 1,600 at 1110km (orange), 400 at 1130km (blue), 375 at 1275km (pink), and 450 at 1325km (green) altitude. Colored lines illustrate ISLs, bright green lines show possible ground-to-satellite links for a ground station~\cite{Kassing2020-yc}.}
    \Description[Overview of Starlink phase one constellation]{The constellation comprises thousands of satellites arranged in shells of evenly-spaced orbits.}
    \label{fig:constellation}
\end{figure}

\Cref{fig:constellation} shows an overview of the satellites in the planned phase \textrm{I} Starlink constellation, which comprises five such shells.
The first shell has 1,584 satellites at altitudes of 550km and is split into 72 planes of 22 satellites each.
Each of these orbital planes is inclined at an angle of 53° to the Earth's equatorial plane.
Additional shells at higher altitudes provide greater areas of coverage at the expense of additional network delay~\cite{Kassing2020-yc,Handley2018-ay}.
The ISLs of such a constellation are likely to be arranged in a \emph{+GRID} pattern, where each satellite keeps a link to its predecessor and successor within its plane as well as to one neighbor each in the two closest adjacent planes.
This allows any two ground stations to connect directly to each other over the satellite network~\cite{Bhattacherjee2019-jz}.

\subsection{Bringing Compute to the LEO Edge}

The edge computing paradigm is the answer to a growing demand and need to process data close to its origin rather than in distant cloud data centers~\cite{paper_bermbach_fog_vision, Zhang2015-cb}.
In fog computing~\cite{paper_bonomi_fog} and the terrestrial MEC architecture~\cite{etsi2022v311}, compute resources are embedded within the access network, i.e., close to clients, the \emph{network edge}.
Here, edge applications run on virtualization infrastructure, which makes them available with high bandwidth and low latency, with lower network costs, and with decreased privacy and security risks~\cite{7469991,paper_pallas_fog4privacy}.

In large LEO satellite communication networks, the network edge is the satellite constellation itself -- the last provider-operated hop before packets reach user equipment.
Uplinks for groups of ground stations converge in a satellite, making it an efficient location for shared computing resources close to clients~\cite{pfandzelter2020edge}.
Possible applications for LEO edge computing include CDN replication to reduce bandwidth usage in the network and access latency for clients, meetup servers for multi-user interaction such as low-latency gaming or video conferencing, or processing space-native data~\cite{Bhosale2020-aa,pfandzelter2020edge,Bhattacherjee2020-kr} -- a detailed overview of the architecture of such an application is given in our case-study in \cref{sec:buoys}.
Given the current skew of cloud and cloudlet data centers being placed close to metropolitan areas, LEO edge computing could help meet the needs of rural areas through the global coverage of LEO satellite networks.

A number of challenges still lie ahead on the path to global LEO edge computing:
First, satellite constellations themselves are still being built out and do not yet provide uninterrupted global coverage.
As a result, there are no system traces from real world LEO satellite constellations.
Second, none of the communication satellites that are already in operation at the moment provide sufficient additional compute resources that could be sold to customers, as satellites are limited by payload restrictions and launch costs.
Although multi-tenant satellites, e.g., the \emph{F6} fractionated satellite~\cite{Brown2009-wz} or the \emph{Tiansuan} satellites~\cite{wangtiansuan}, have been researched, to the best of our knowledge, no operational compute platform for LEO satellite networks exists at the moment.
Third, the deployment and operation of such resources in LEO introduce additional engineering challenges in power consumption, waste-heat management, or radiation hardening~\cite{Nedeau1998-fj,Bhattacherjee2020-kr,stauning2004detection}.
Fourth, how to abstract from the highly mobile and unique infrastructure using LEO edge computing platforms is still actively being researched, e.g.,~\cite{Bhosale2020-aa,Pfandzelter2021-dp}.

\subsection{Open Challenges in Building and Testing LEO Edge Software Systems on Earth}

A key part of systems research for the LEO edge is the evaluation of software systems in a realistic environment, e.g., through functional testing and benchmarking.
While algorithms can be evaluated in simulation, studying the behavior of concrete software systems requires a realistic runtime environment.
We believe that enabling researchers and practitioners to quickly and affordably run virtual LEO edge computing infrastructure in the cloud will accelerate both systems research and the actual development and deployment of infrastructure as operators gauge industry and research interest.
Yet, building such testbeds for large-scale, highly dynamic infrastructure requires addressing three specific research challenges:
How can we ensure that a testbed offers an accurate representation of real LEO constellations?
Which abstractions are necessary in the design of a testbed so that arbitrary software systems can be deployed on it?
Finally, how can we achieve this with cost-efficiency at scale?

\paragraph*{Accurate Emulation of LEO Constellations}

A fundamental difference of the LEO edge compared to terrestrial edge or cloud computing is the high mobility of satellites in relation to Earth that result from their low orbits.
A complete LEO edge infrastructure is constantly evolving, with network connections and characteristics between servers changing by the second.
Furthermore, ground stations also frequently switch their uplink to their  closest satellite as a result of this mobility.
All applications and platforms developed for the LEO edge need to consider this and proactively replicate data and services.
These characteristics of the network influence the performance of edge applications, which are often latency or bandwidth sensitive, and must thus also be part of a LEO edge computing testbed.
This should be based on detailed models of Earth and space so that hand-off and data migration techniques can be evaluated as accurately as possible.

Satellite servers will likely use commercial, off-the-shelf compute hardware, e.g., as HPE does with their \emph{Spaceborne Computers} onboard the International Space Station~\cite{noauthor_2021-jm,noauthor_2019-ui}.
Although the effects of the Van Allen radiation belts are negligible in lower orbits, servers will be subject to some single event upsets caused by intermittent galactic cosmic rays that can normally be absorbed by Earth's atmosphere and magnetic field~\cite{Koontz2018-pb}.
HPE has shown that this can be remediated with standard software and hardware mechanisms, yet at the cost of temporary performance degradation or full shutdowns.
Such failure will impact any software running on a satellite and developers will want to test their applications against these scenarios.
Especially service orchestrators on top of LEO edge infrastructure need to adapt and react quickly.

\paragraph*{Support for Arbitrary Software Systems}

A LEO edge computing testbed should enable development and evaluation of any kind of LEO edge software and should not be restricted to certain programming languages, frameworks, or deployment models.
The emulator should thus accurately reflect blank-slate servers, especially as we still face unknowns regarding middleware platforms and applications.

When considering the development of platforms, great care should be taken to ensure that a virtual testbed does not impose restrictions on the kind of technologies that can be evaluated within, e.g., a container-based approach where each satellite server is emulated with a single container would impose limitations on testing container-based platforms.

\paragraph*{Cost-Efficient Scalability}

Large LEO satellite constellations can comprise tens of thousands of satellites and serve millions of clients.
Building and scaling applications and platforms for such infrastructure is a difficult challenge and a LEO edge testbed should be able to serve as a way to perform scalability evaluation.
The testbed must thus also scale for large constellations and provide the means to emulate such infrastructure.

However, users should not be expected to also provide tens of thousands of physical computers for such an emulation.
Rather, the testbed should be cost-efficient in a way where more than one satellite server can be emulated on one host.
Further, it should also let the user evaluate only a part of the complete constellation, e.g., to support rapid prototyping and testing of a constellation subset over a certain geographical area or to save experiment costs.

\subsection{Related Work}

In the edge computing, where applications are widely distributed and physical infrastructure is often inaccessible, researchers commonly rely on virtual testbeds, network simulators, and edge simulators for testing.

Testbed tools such as \emph{Fogbed}~\cite{Coutinho2018-wa}, \emph{EmuFog}~\cite{Mayer2017-dt}, or \emph{MockFog}~\cite{Hasenburg2019-er,9411706} allow users to create and manipulate virtual infrastructure that mimics a distributed, heterogeneous edge-cloud continuum.
Fogbed and EmuFog, however, do not support an emulation of highly dynamic network topologies as required for large LEO satellite constellations.
Furthermore, applications are deployed as Docker containers, which makes it unsuitable for evaluating novel platforms as it limits supported software.
MockFog supports dynamic network changes and uses a dedicated cloud virtual machine for each compute node, which imposes fewer restrictions on the kinds of software it supports, yet such flexibility comes at a price:
With a dedicated cloud virtual machine for each satellite server, we cannot achieve a cost-efficient emulation for large LEO constellations.
Similar concerns can also be raised for further IoT and edge computing testbed tooling~\cite{hashemian2020contention,10.1145/3365871.3365897,behnke2019hector,balasubramanian2014rapid,Eisele2017-yx}.

On the other hand, edge simulators such as \emph{iFogSim}~\cite{Gupta2017-jx} or \emph{FogExplorer}~\cite{paper_hasenburg_supporting_2018,paper_hasenburg_fogexplorer_2018} are more cost-efficient and allow evaluating larger topologies over a longer time period.
These could be extended to also simulate the ``movement'' of (LEO) edge nodes, yet would still not allow users to run their actual application in experiments, thus limiting accuracy of results to the assumptions of the underlying simulation model.

Finally, network simulators such as \emph{ns--3}~\cite{ns3} let users explore network-level effects such as congestion or routing in large-scale networks.
Kassing et al.~have presented \emph{Hypatia}~\cite{Kassing2020-yc}, a network simulator for large LEO satellite constellations based on~\emph{ns-3} that allows researchers to measure LEO satellite network characteristics on a packet-level.
Similarly, \emph{SILLEO-SCNS}~\cite{Kempton2021-lw}, which \textsc{Celestial}'s Constellation Calculation is based on, also facilitates studying and visualizing such networks.
While important in their own right, these network simulators target a different use case, namely the evaluation of network measurement, and cannot be used to evaluate software systems.

In fact, to the best of our knowledge, there is no testbed tooling which allows users to evaluate LEO edge software and, consequently, systems researchers have until now not been able to study the impact of the LEO edge environment on real applications and platforms.

\section{Emulating the LEO Edge with \textsc{Celestial}}
\label{sec:systemdesign}

\begin{figure}
    \centering
    \includegraphics[width=\columnwidth]{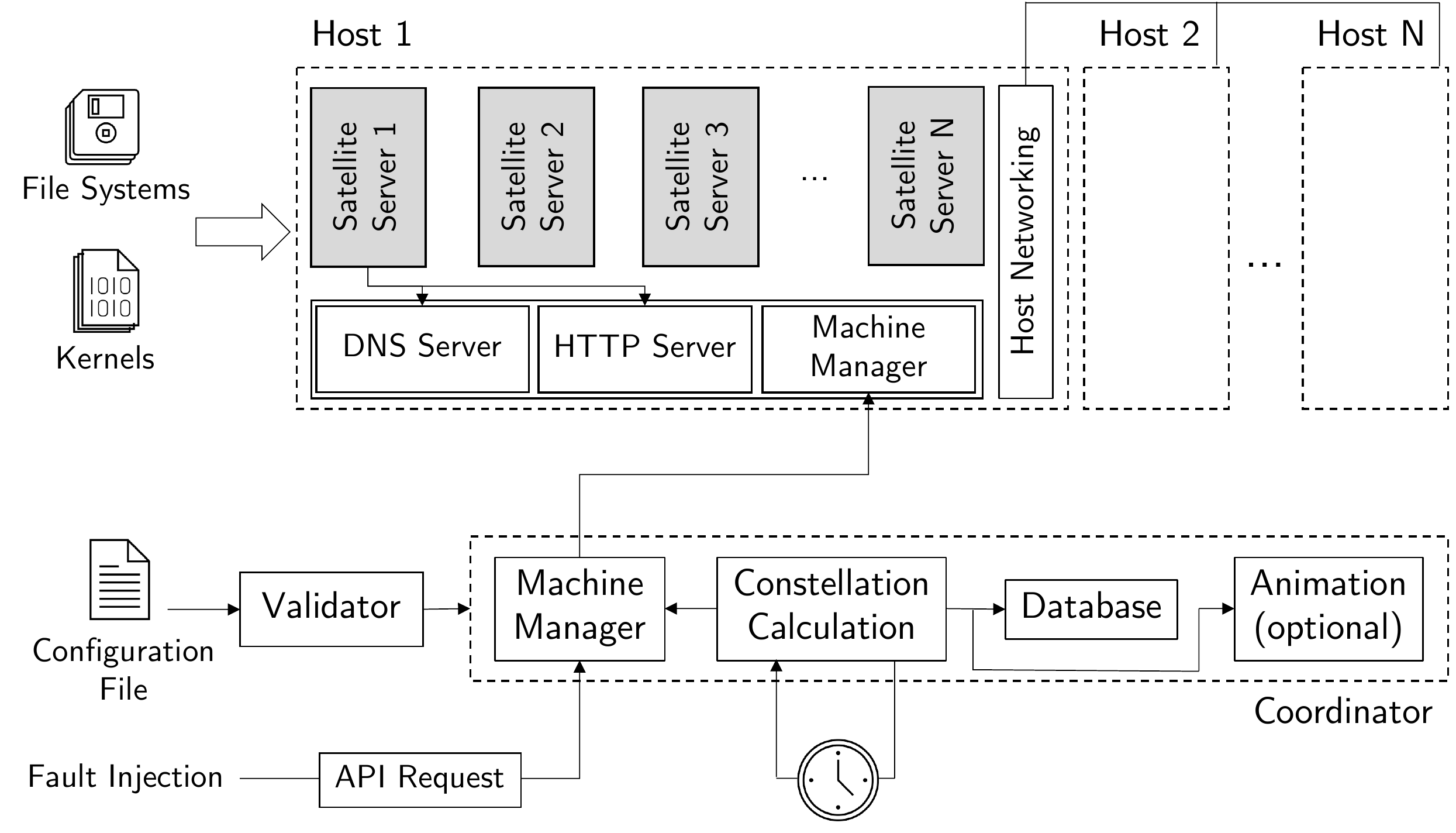}
    \caption{\textsc{Celestial}'s coordinator calculates satellite positions and updates machines and network links on \textsc{Celestial}'s hosts.}
    \Description[Architecture of \textsc{Celestial}]{\textsc{Celestial}'s coordinator calculates satellite positions and updates machines and network links on \textsc{Celestial}'s host.}
    \label{fig:architecture}
\end{figure}

We present \textsc{Celestial}, a novel emulation tool for edge computing on large LEO satellite constellations.
An overview of its architecture is shown in \cref{fig:architecture}.
\textsc{Celestial} can run on an arbitrary number of standard Linux servers, e.g., in the cloud, and comprises two main components:
A central coordinator computes satellite orbital paths and networking characteristics.
This information is sent to \textsc{Celestial} servers that host a microVM for each satellite server and ground station.
\textsc{Celestial} servers also manipulate network connections between microVMs to accurately reflect satellite movement, available links, as well as their delays and bandwidth.
In this section, we describe how we address the open research challenges with \textsc{Celestial}.

\subsection{Accurate Emulation of LEO Constellations}

At the heart of \textsc{Celestial}, the \emph{Constellation Calculation} component updates the state of the satellite network periodically, including the positions of satellites and ground stations, network link distance and delays, and shortest paths between nodes.
This is based in large parts on the \emph{SILLEO-SCNS} network simulator~\cite{Kempton2021-lw} for large LEO satellite constellations, which we extend with support for the \emph{SGP4} simplified perturbations models.
Additionally, we use more efficient implementations of Dijkstra's algorithm~\cite{dijkstra1959note} and the Floyd-Warshall algorithm~\cite{10.1145/367766.368168} to calculate the shortest network paths within the constellation and their end-to-end latency.

SGP4 is the state-of-the-art in calculating the positions of satellites and takes perturbations caused by atmospheric drag, the Earth's shape, and gravitational effects from Moon or Sun into account~\cite{hoots1980models}.
The model input parameters can be obtained from the database of NORAD two-line element sets (\emph{TLE})\footnote{\url{https://celestrak.com/NORAD/elements/}} for satellites already in orbit or calculated based on simple parameters such as a satellite shell's inclination and altitude.
To limit side effects and ensure repeatable testing, all parameters are passed within a single configuration file.
This includes network parameters, such as ISL bandwidth, compute parameters that describe the allocated resources for satellite and ground station servers, orbital parameters for satellite shells to support different kinds of constellations, and ground station locations.

\textsc{Celestial} then uses satellite and ground station positions to calculate ground station uplinks and the satellite network topology.
Here, ISL connectivity depends on the line of sight between two adjacent satellites, e.g., if a possible laser link drops below a certain altitude, the Earth's atmosphere may refract that laser, causing an intermittent loss of connectivity.
Similarly, ground stations can only communicate with satellites that are above a configurable minimum elevation above the horizon.
In our tests, these calculations could be completed within one second even on a standard laptop.
The results are then transferred to the hosts, where network delays and bandwidth constraints between the satellite servers are emulated using \texttt{tc}\footnote{\url{https://man7.org/linux/man-pages/man8/tc.8.html}} and \texttt{tc-netem}\footnote{\url{https://man7.org/linux/man-pages/man8/tc-netem.8.html}}.
Emulated network delays are injected with a 0.1ms accuracy and any latency between hosts is taken into account, yet this only works if this latency is low enough, e.g., hosts are located in the same datacenter.
Additionally, emulated servers can still reach the Internet through the host, e.g., to store experiment data in some central location.

We isolate microVMs in dedicated \texttt{cgroups} to gain more finely grained control over the CPU cycles a server process is allowed to use, making the emulation of severely constrained satellite servers possible.
Through an API, users can change machine parameters at runtime and even terminate and reboot machines to model faults, e.g., caused by radiation.

\textsc{Celestial} also provides an optional animation component that visualizes the state of the constellation during the emulation run, e.g., \cref{fig:constellation} was generated by this component.
We believe that this can be a great help in understanding the characteristics of satellite mobility and networking as well as the effects on their software systems, especially for developers who are new to satellite networks.

In \textsc{Celestial}, we take all effects into account that are currently conjectured to have a significant impact on software systems running on the LEO edge, namely dynamic network delays, bandwidth restrictions, constrained compute and storage resources, and service interruption through environmental effects such as radiation~\cite{Bhattacherjee2020-kr,Pfandzelter2021-dp}.
When additional effects are studied and traces from real LEO edge deployments become available, researchers will be able to quickly incorporate the necessary changes in \textsc{Celestial}:
First, the core Constellation Calculation component that dictates the effects that are emulated in the rest of the system comprises only 971 lines of Python source code that can be easily customized and extended.
The separation from Machine Managers allows the resulting networking and machine parameters to be sent to host machines without modification.
Second, \texttt{tc-netem} offers advanced network emulation features that are not currently used in \textsc{Celestial}, such as packet loss or duplication, delay distributions, packet corruption, or packet reordering.
If such characteristics will be useful for LEO edge testbeds in the future, they can be added with only small changes to the \textsc{Celestial} codebase.
Third, we forego the use of complex dependencies such as orchestrators (e.g., Kubernetes or Mesos) and overlay network tooling (e.g., Flannel or Calico) to reduce the complexity of \textsc{Celestial}.
While we rely on some of their underlying technologies, such as the \texttt{firecracker-containerd} plugin\footnote{\url{https://github.com/firecracker-microvm/firecracker-containerd}}, not including such complex software simplifies prototyping new features and reduces maintenance efforts of our implementation, especially considering the required changes to such dependencies.

\subsection{Support for Arbitrary Software Systems}

All satellite and ground station servers are emulated with \emph{Firecracker}~\cite{Agache2020-ug} microVMs that run on the \textsc{Celestial} hosts, managed by the \emph{Machine Manager} components.
Firecracker microVMs provide a sub-second boot time and support for microVM suspensions, and can use configurable Linux kernels and root filesystems.
\textsc{Celestial} thus gives users the control over kernel features, installed software, and programming and deployment models for their applications.
The process of compiling filesystems, starting (cloud) hosts, and uploading the necessary files can be automated using common orchestration tools such as \emph{Ansible}\footnote{\url{https://www.ansible.com/}}, yet it is highly user-specific, so we do not include it in \textsc{Celestial}.
Crucially, satellite servers are thus provided as a blank-slate and users may even set up container technologies such as Docker within their emulated servers.
This is of considerable importance for systems researchers, as it also facilitates the evaluation of container-based application orchestration using tools such as Kubernetes or FaaS platforms~\cite{Pfandzelter2020-kw,Bhosale2020-aa}.

To aid the development of applications and platforms on \textsc{Celestial}, it includes two additional components:
First, each \textsc{Celestial} host provides a local DNS server that can resolve microVM network addresses with a custom DNS record.
Applications can simply query the \texttt{A} records for, e.g., \texttt{878.0.celestial} to get the network addresses of satellite 878 in the first shell.
Applications thus do not have to be aware of the underlying IP address space calculation for \textsc{Celestial}'s virtual network interfaces.

Second, \textsc{Celestial} hosts run an HTTP server that provides information on satellite positions, network paths between satellites, constellation information, and more to the emulated satellite servers.
This information is sourced from a central database on the \textsc{Celestial} coordinator that is updated by the Constellation Calculation.
Application developers can leverage this API to quickly test their applications on different kinds of LEO constellations without having to implement a custom model of satellite movement and network behavior.
In a real LEO edge computing scenario, we expect such information to be available from a central source provided by the satellite network operator or from public sources such as a TLE database.
However, users are free to use only a subset or none of these APIs, e.g., when testing their own models.

\subsection{Cost-Efficient Scalability}

To emulate arbitrarily large LEO and complex LEO satellite constellations with thousands of satellite servers, \textsc{Celestial} supports horizontal scale-out across many hosts over which microVMs are distributed.
For this, users may simply instantiate necessary cloud infrastructure that can be terminated once experiments and tests are completed.
\textsc{Celestial} automatically creates an overlay network using \emph{WireGuard}~\cite{donenfeld2017wireguard}, thus connecting hosts and offering routing between microVMs.
Further, the use of microVMs allows for high over-provisioning as well as collocation of many emulated satellite servers on few physical servers.
In \textsc{Celestial}, we use the fact that all satellite servers are identical to our advantage and de-duplicate microVM root filesystems using a common immutable disk image in addition to an overlay for each microVM, allowing us to save on storage space and improve performance.

\textsc{Celestial} can emulate large and complex LEO satellite constellations with thousands of satellite servers, yet not every test or evaluation requires emulating each individual satellite server at the same time:
As an optional feature, we introduce a configurable bounding box, a geographical area on Earth to which emulated satellite servers are limited.
As satellites are mobile, they can quickly move in and out of this bounding box and may only stay relevant for evaluation for a few minutes.
To free up resources, satellite microVMs are suspended when they move out of the bounding box and re-activated when they come back into it.
The underlying idea is that in edge computing, clients will want to use edge servers that are in their proximity, and distant edge servers will thus not need emulation.
To give an example, in \cref{sec:videoconferencing}, we use the bounding box to only emulate satellites over North Africa, where our clients are located, to save resources.
\textsc{Celestial} also helps the user configure their bounding box in a manner that makes sure that available resources meet the demand from the emulation based on per-microVM resources and bounding box area.
Note that this bounding box does not affect network path calculation, as the shortest network path between two ground stations may not follow the line of sight.
This allows algorithms, applications, and platforms for large LEO satellite constellations to be tested cost-efficiently on a minimal subset of servers before moving to an emulation of the entire constellation.

\section{Deploying an Edge Application on \textsc{Celestial}}
\label{sec:videoconferencing}

\begin{figure}
    \centering
    \includegraphics[width=\columnwidth]{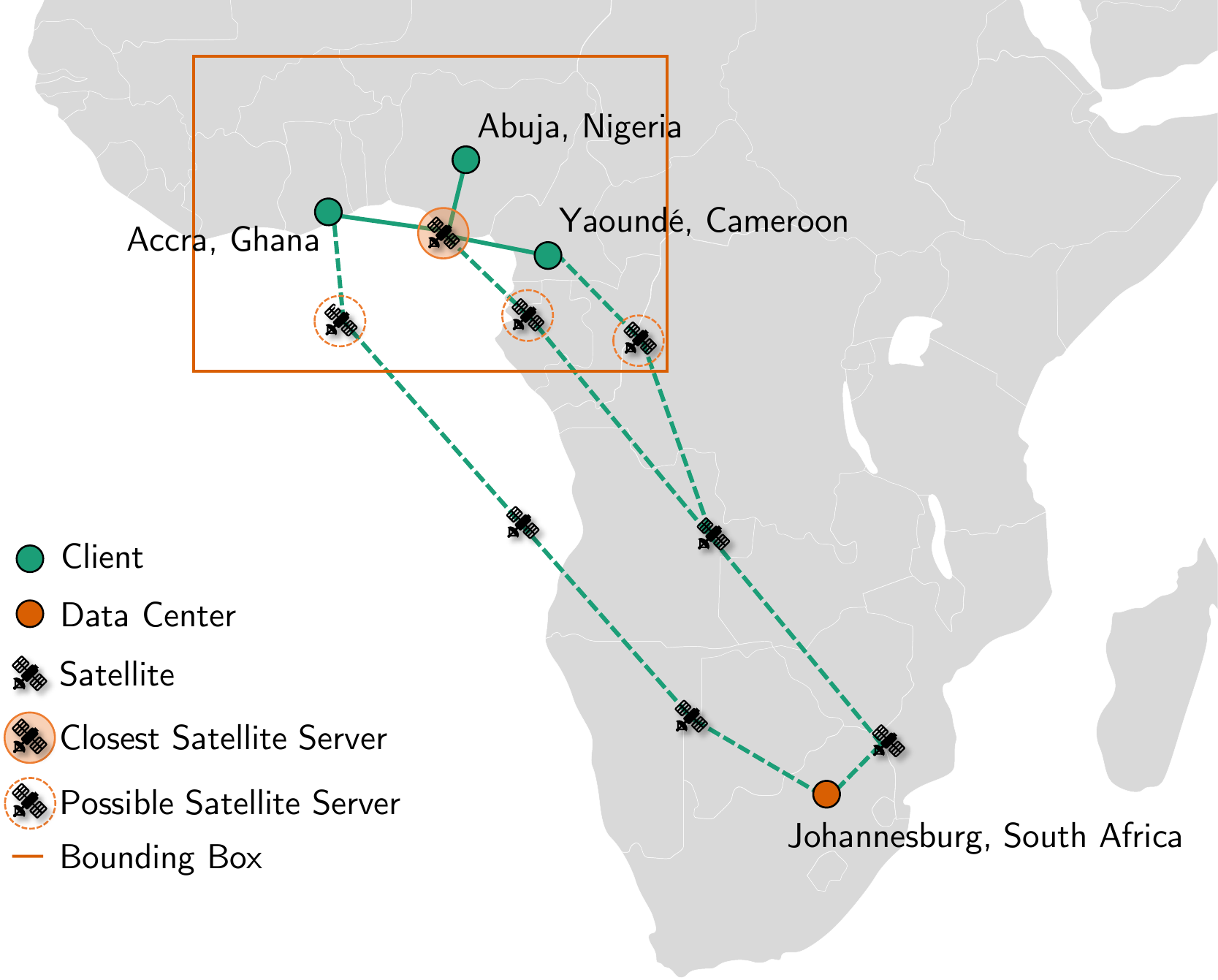}
    \caption{Three clients in Accra, Ghana; Abuja, Nigeria; and Yaoundé, Cameroon, require a common meetup-server for an interactive application. This meetup server may be located in a satellite server or the nearest cloud datacenter in Johannesburg, South Africa. Dashed lines indicate the additional hops needed to reach the cloud datacenter~\cite{Bhattacherjee2020-kr}.}
    \Description[Clients in Accra, Abuja, and Yaoundé use an interactive application]{Using the cloud datacenter in Johannesburg, South Africa, for their meetup server incurs additional network distance.}
    \label{fig:example}
\end{figure}

To evaluate \textsc{Celestial}'s performance, accuracy, and scalability, we first deploy an example LEO edge application, namely a multi-user interaction between users in West Africa as presented by Bhattacherjee et al.~\cite{Bhattacherjee2020-kr}.
In this example, which we illustrate in \cref{fig:example}, three users located in Accra, Ghana; Abuja, Nigeria; and Yaoundé, Cameroon require a common meetup-server for their application.
While their nearest available cloud data center is located in Johannesburg, South Africa, using a satellite server reduces the RTT for the most distant of the three users from 46ms to only 16ms over SpaceX' phase \textrm{I} Starlink network.

We implement this as a WebRTC video conference where each participant sends high-definition video at 2.6Mb/s and receives a video stream from the other participants.
An intermediary bridge server, our meetup service, duplicates each user's stream for all other users rather than each user sending multiple copies of their stream at the same time, which would put a considerable strain on their bandwidth.
We can now compare two scenarios:
First, we run the video bridge on the datacenter, which we assume to have an antenna to access the satellite network~\cite{AzureSpace-ro}, which is the best-case scenario for latency to our clients.
Second, we deploy a small tracking service to this datacenter that periodically checks the satellites in reach of our clients and instructs them to use the optimal satellite server based on combined latency as a video bridge server.
As the video bridge for our real-time video conferencing use-case can be considered stateless, we do not take any migration costs into account.
We discuss the issue of state management on the LEO edge in \cref{subsec:future_leo_work}.

\begin{figure*}
    \centering
    \begin{subfigure}{0.31\textwidth}
        \centering
        \includegraphics[width=\linewidth]{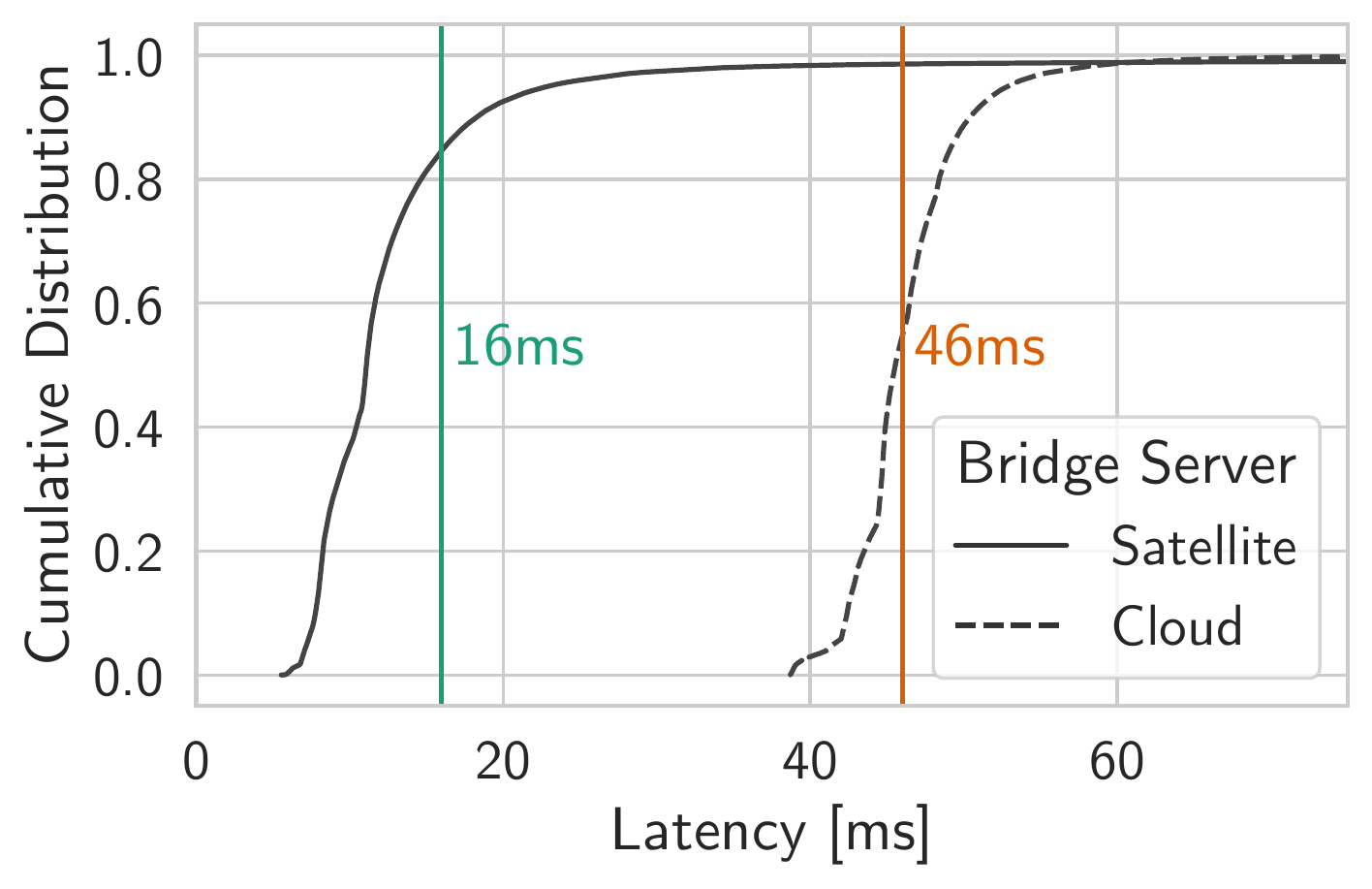}
        \caption{Accra to Abuja}
        \label{fig:cdf-12}
    \end{subfigure}%
    \hfill
    \begin{subfigure}{0.31\textwidth}
        \centering
        \includegraphics[width=\linewidth]{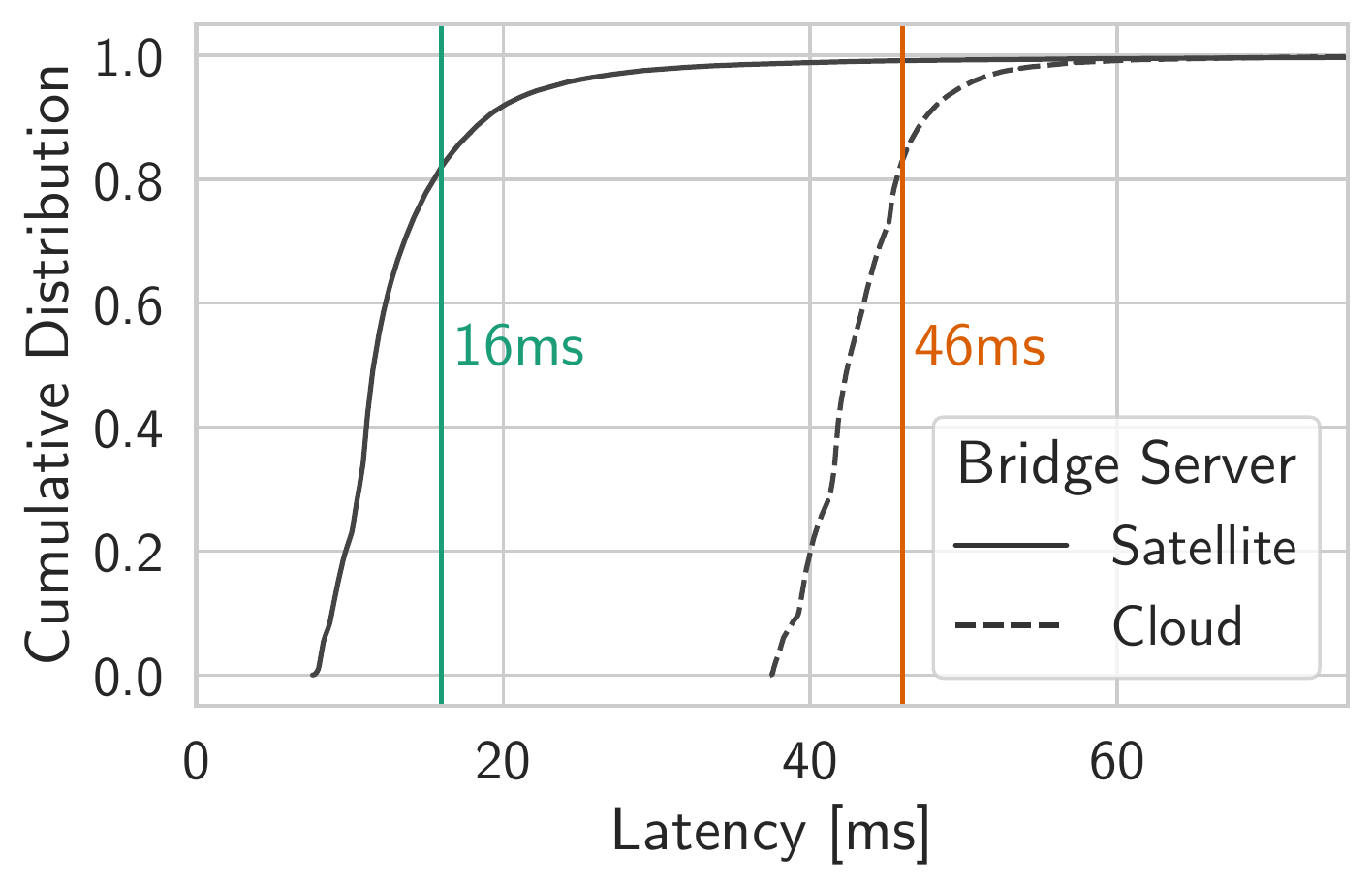}
        \caption{Accra to Yaoundé}
        \label{fig:cdf-13}
    \end{subfigure}%
    \hfill
    \begin{subfigure}{0.31\textwidth}
        \centering
        \includegraphics[width=\linewidth]{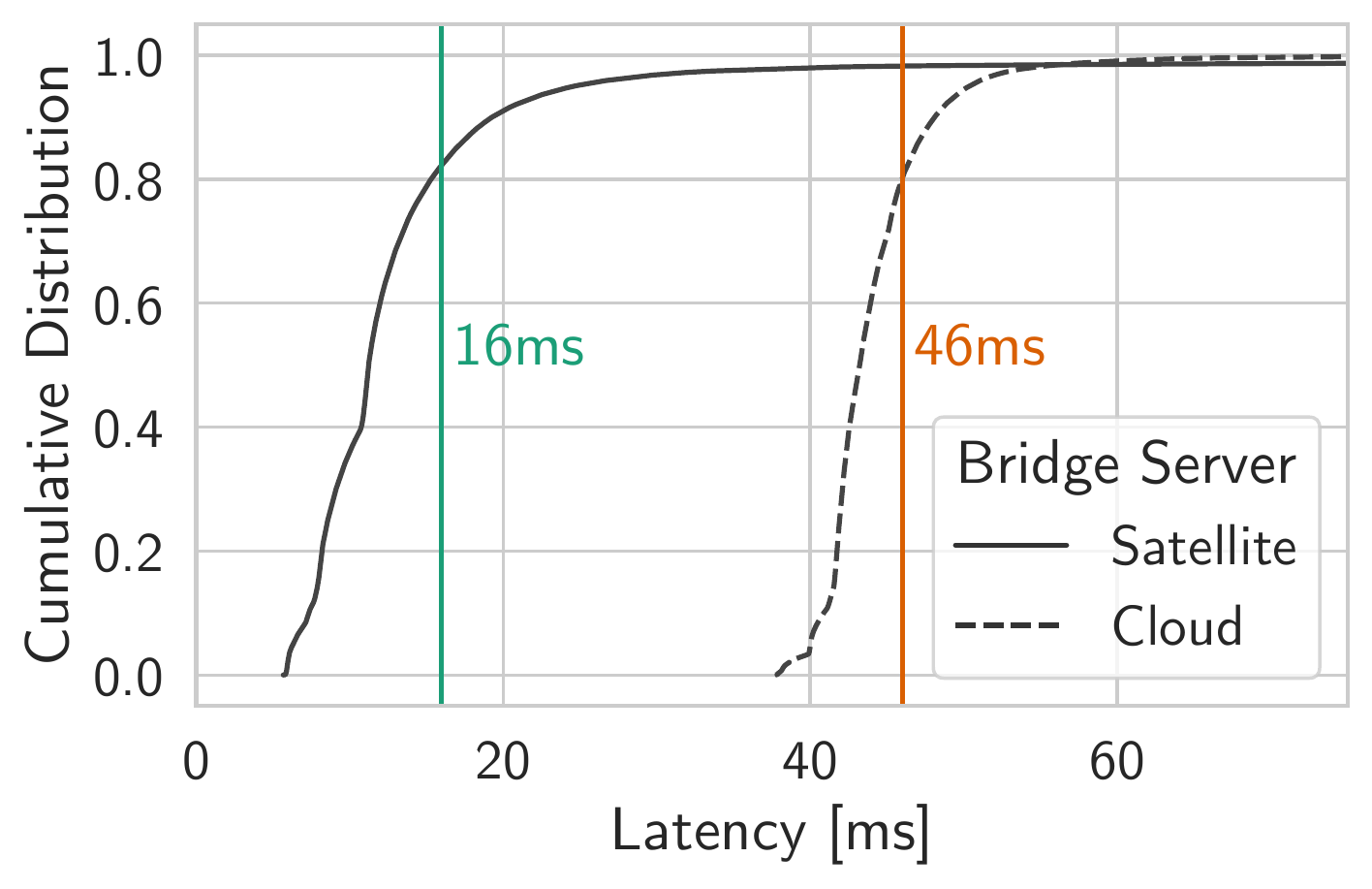}
        \caption{Abuja to Yaoundé}
        \label{fig:cdf-23}
    \end{subfigure}
    \caption{Our results show that the satellite servers can offer better QoS over the cloud data center, with only up to 15ms end-to-end latency for our users for 80\% of the duration of their video conference.}
    \Description[Cumulative distribution of end-to-end latency for our user-pairs.]{Results show a significant improvement in end-to-end latency with satellite servers over a cloud server. 80\% latency drops from around 80ms to 15ms.}
    \label{fig:cdfs}
\end{figure*}

\subsection{Experiment Setup}

We assume both laser propagation in a vacuum and propagation of ground-to-satellite RF links at speed of light $c$~\cite{Smith-Rose1950-ix,Handley2018-ay} and a 10Gb/s bandwidth connection for ISLs and radio links.
We allocate four CPU cores and 4GB memory for each client and the tracking service, while each satellite server as well as the cloud video bridge are allocated two CPU cores and 512MB memory.
To limit resource usage and the number of satellite servers the tracking service must consider, we draw a bounding box as shown in \cref{fig:example}.
To minimize the impact of clock drift in delay measurements between our clients, we schedule them to run on the same \textsc{Celestial} host and instruct them to use a shared \emph{PTP} clock.
In a preliminary baseline evaluation, we find that our clients and bridge server incur a 1.37ms median processing delay (3.86ms standard deviation) that we must take into account when comparing measured end-to-end latency and expected network distance.
We expect our client measurement software, packet duplication, packet forwarding, and clock drift to cause this jitter.

Our experiments are conducted on three Google Cloud Platform \texttt{N2-highcpu} instances with 32 cores and 32GB memory each, placed in the \emph{europe-west3-c} zone, connected in a private virtual network, and using an installation of Ubuntu 18.04.
We measure a network delay of 0.2ms between those machines, yet this delay is already taken into account by \textsc{Celestial} when emulating microVM network distance.
Although larger instances are available that may allow us to host our testbed on only a single cloud server, our goal is to show the horizontal scalability of \textsc{Celestial}.
While \textsc{Celestial} estimates 137 required CPU cores given satellite density and bounding box size, we use only 96 CPU cores to test its over-provisioning capabilities.
As our application actually uses resources on only one satellite server at a time, we expect only a small load on our \textsc{Celestial} servers.
In addition, we run our \textsc{Celestial} Coordinator on a GCP \texttt{C2} instance with 16 cores and 64GB memory and choose an update interval of two seconds.
Each experiment is ten minutes long and repeated three times to validate its reproducibility.

\subsection{Results}

We start by giving an overview of the results of our experiments by presenting the results of a randomly selected run.
We can then validate the accuracy of the network conditions in our \textsc{Celestial} testbed by comparing it to the simulated network, analyze the reproducibility of our experiments on \textsc{Celestial}, and finally check resource efficiency\footnote{We include additional results in a tech report on \textsc{Celestial}~\cite{techreport_pfandzelter_2022}.}.

We show the cumulative distributions of measured end-to-end latency between client pairs in \cref{fig:cdfs}.
For at least 80\% of the duration of the video conference, end-to-end latency is below this maximum RTT of 16ms for the satellite servers and 46ms for the cloud server, confirming our expectation that satellite servers can provide a considerable QoS improvement for clients in latency-sensitive applications.

While available bandwidth is an unlikely factor for increased latency in the remaining 20\% given the small size of our video stream, processing delay may add several milliseconds.
Additionally, we cannot assume an optimal server to be selected at all times as, in some instances, the satellite constellation may change noticeably during the five-second update interval used in our application, e.g., with a particular uplink satellite becoming unavailable.

We further observe that only satellites in the two shells with the lowest altitude and highest density are ever selected as satellite servers, as they are more likely to have an optimal connection to all three clients at the same time.
In deploying LEO edge infrastructure, it might thus make sense to start with the densest and lowest altitude shell.

\paragraph{Accuracy}

\begin{figure}
    \centering
    \includegraphics[width=0.95\columnwidth]{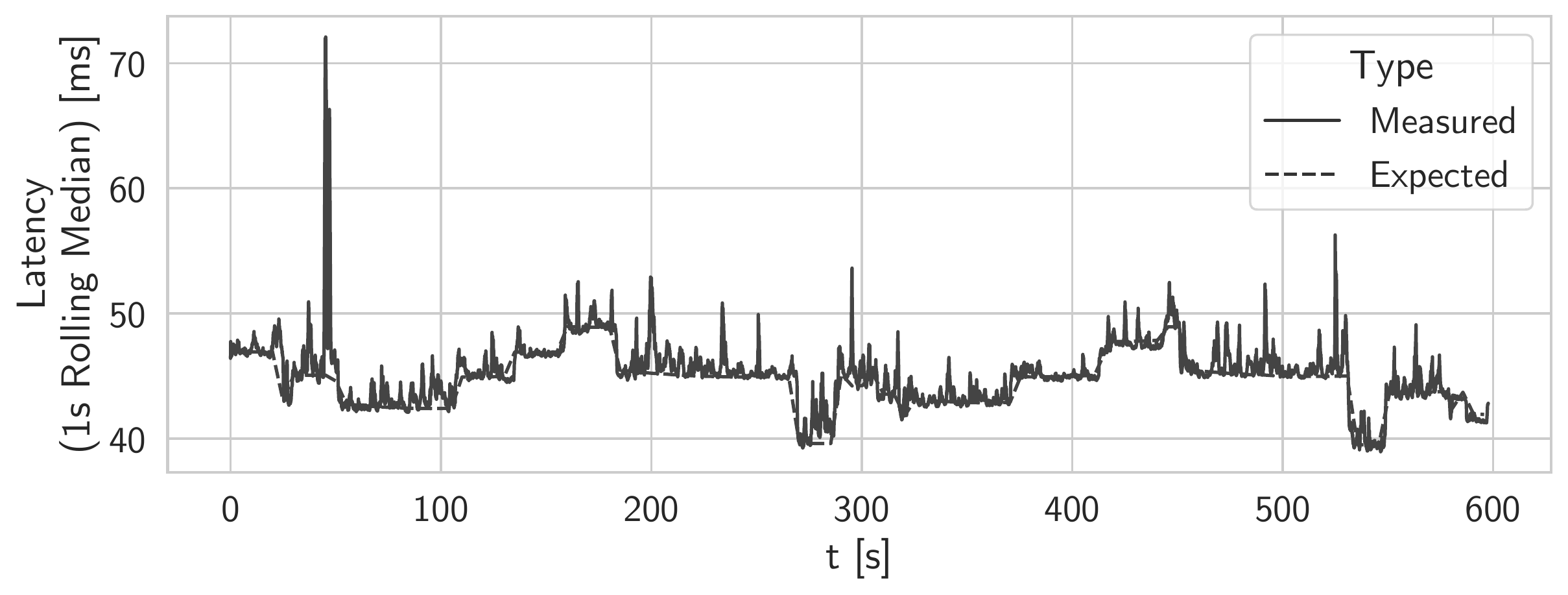}
    \caption{Measured and expected end-to-end latency from Abuja to Accra using the Johannesburg cloud datacenter. The expected value includes both simulated network distance and median processing delay of 1.37ms.}
    \Description[expected and measured end-to-end latency from Abuja to Accra]{Both expected and measured end-to-end latency follow the same curve.}
    \label{fig:accuracy}
\end{figure}

To investigate whether the measured latency over time is accurate, we compare \emph{expected} network latency as calculated by our tracking server, which includes the 1.37ms median processing delay.
As an example, we show this comparison for the path from Abuja to Accra using the cloud server over the course of one evaluation in \cref{fig:accuracy}.

Both curves follow the same general trend and changes in calculated network latency are reflected in measured end-to-end latency.
Again, we note that the tracking server only chooses a new server at the coarse interval of five seconds and intermittent changes in infrastructure that can cause spikes in end-to-end latency are thus not reproduced in expected network distance.
Additionally, the jitter in our measurements is likely caused by processing as we observe it in our baseline measurements as well.
We thus conclude that simulated network distances are accurately reflected in the emulated testbed.

\paragraph{Reproducibility}

\begin{figure}
    \centering
    \includegraphics[width=0.95\columnwidth]{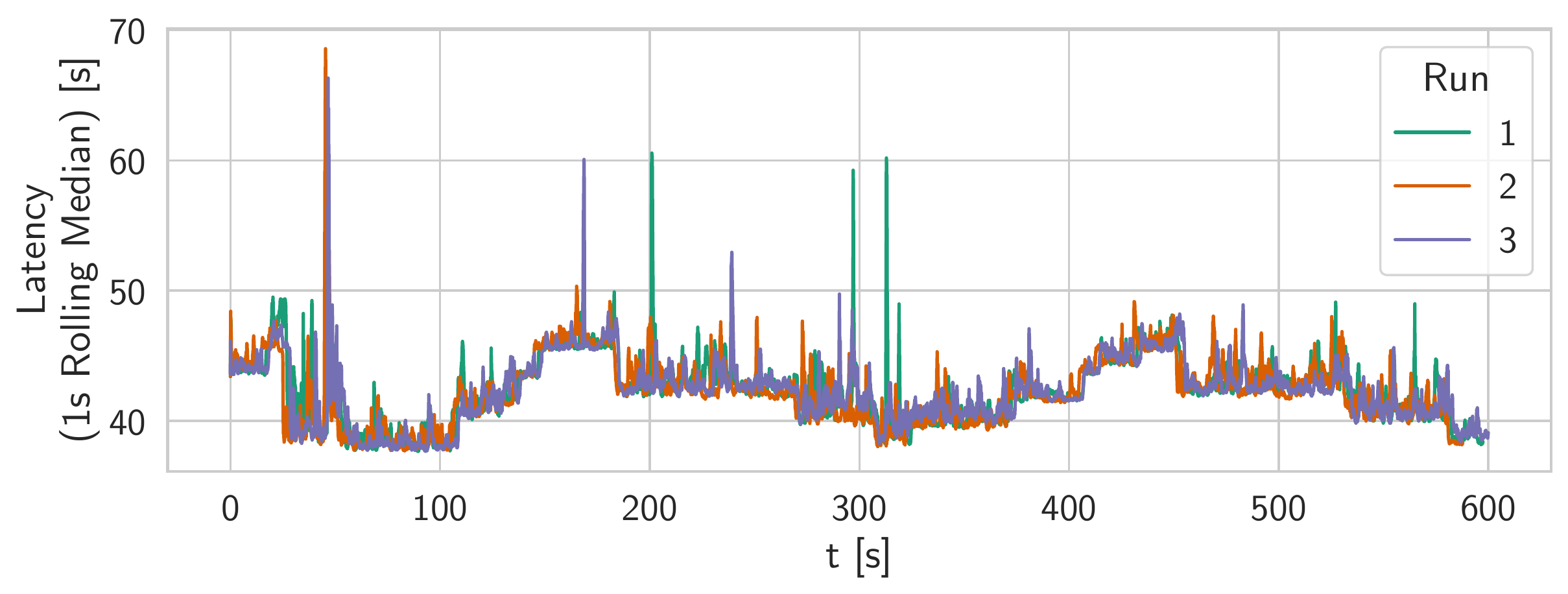}
    \caption{Measured end-to-end latency from Yaoundé to Abuja using the Johannesburg cloud datacenter across three repetitions of the experiment}
    \Description[measured end-to-end latency from Yaoundé to Abuja across three repetitions of the experiment]{Measured results for the experiment across three repetitions show the same trends and reflect repeatable changes in communication latency.}
    \label{fig:reproducibility}
\end{figure}

Furthermore, we can investigate whether our results are reproducible by looking at the results from the three repetitions we carry out for each experiment.
As an example, we plot the measured end-to-end latency from Yaoundé to Abuja using the cloud server across the three repetitions in \cref{fig:reproducibility}.

We can observe that results for all three runs follow the same trends and even the spike in measured latency after the first minute of the experiment can be reproduced.
Since users can provide an arbitrary but firm starting point for their testbed emulation, \textsc{Celestial} offers a repeatable environment that enables reproducible tests and benchmarks.

\paragraph{Efficiency}

\begin{figure}
    \centering
    \includegraphics[width=0.95\columnwidth]{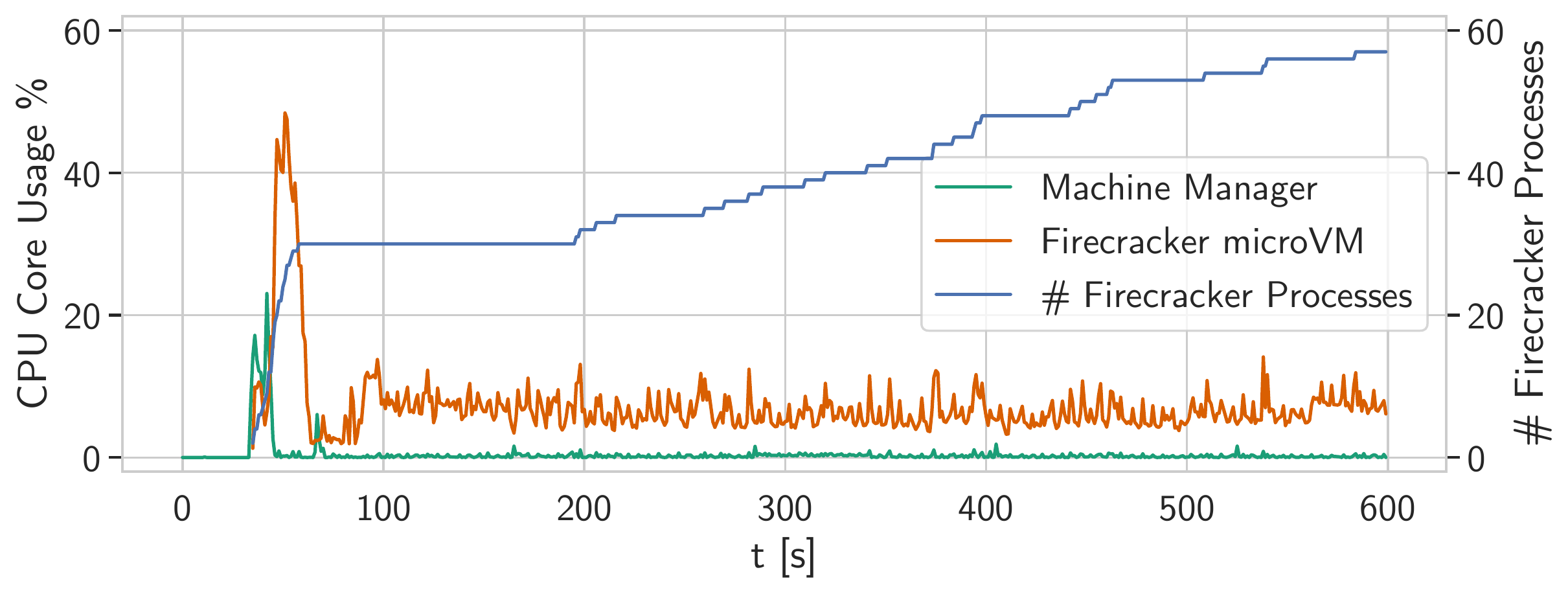}
    \caption{CPU usage on one \textsc{Celestial} host over the course of one experiment. In total, 32 CPU cores are available on the host.}
    \Description[CPU usage over the course of one experiment]{Even with the number of Firecracker processes increasing, the host's CPU usage remains at only 10\%}
    \label{fig:cpu_usage}
\end{figure}

\begin{figure}
    \centering
    \includegraphics[width=0.95\columnwidth]{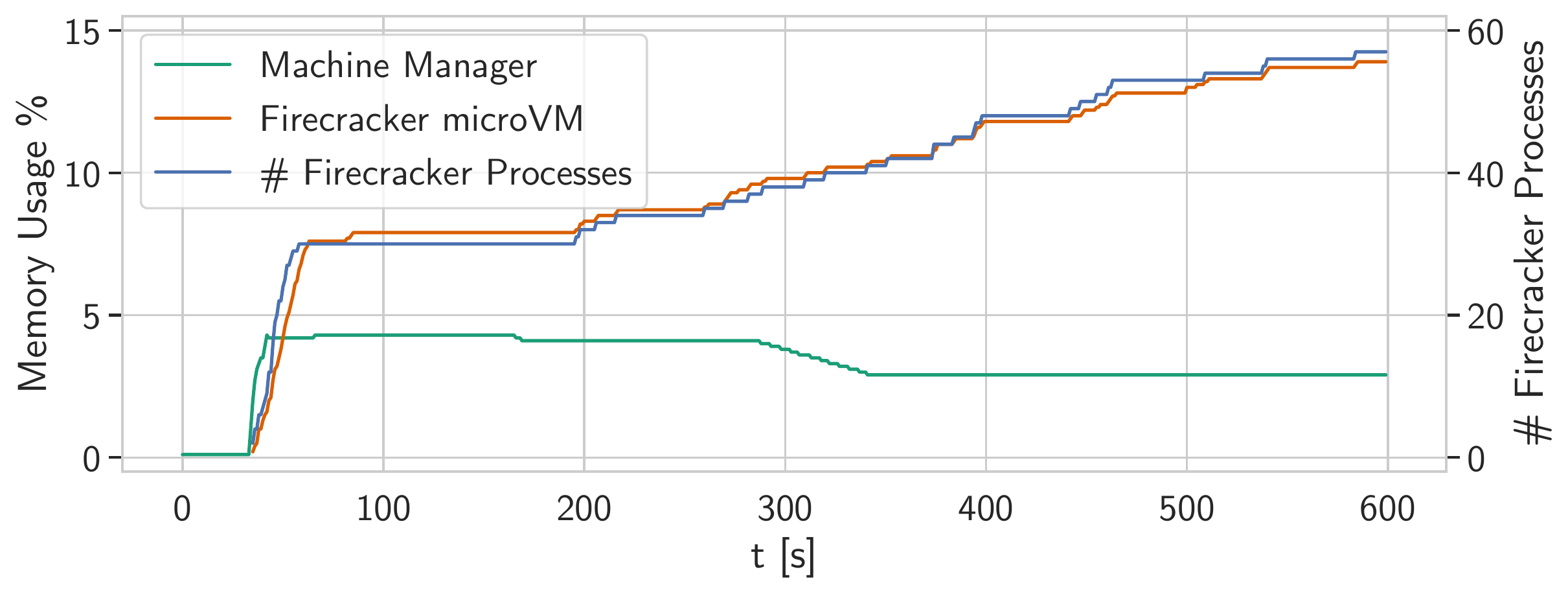}
    \caption{Memory usage on one \textsc{Celestial} host over the course of one experiment. In total, the host has 32GB available memory.}
    \Description[Memory usage over the course of one experiment]{\textsc{Celestial}'s Machine Manager uses only little memory. Firecracker microVMs keep the memory reserved by their \texttt{virtio} memory devices even when suspended.}
    \label{fig:mem_usage}
\end{figure}

Finally, in a separate test we trace the CPU and memory usage on our \textsc{Celestial} hosts for a glimpse into the resource efficiency of our testbed.
Specifically, we look at the host under the highest load, which is that on which all our clients run for accurate time synchronization, in addition to a third of all satellite servers.
We show its CPU and memory utilization in \cref{fig:cpu_usage,fig:mem_usage}, respectively.

At the beginning, we see a spike in CPU usage caused by \textsc{Celestial}'s Machine Manager as it starts setting up the host and network environment, followed by an even larger spike as Firecracker microVMs are starting to boot up at the beginning of the emulation run.
CPU usage then decreases to below 5\% as the clients prepare for the experiment, e.g., by synchronizing clocks and reading the workload traces.
Then, during the experiment, we see total CPU usage from microVMs on the order of 10\%.
Considering that three clients and one tracking server with four allocated CPU cores each, and one active bridge server with two allocated cores run a demanding workload, and at least 25 additional satellite servers idle, we can conclude that \textsc{Celestial} is efficient on CPU resources as it can benefit from over-provisioning.
Additionally, note that \textsc{Celestial}'s Machine Manager itself consumes few CPU resources after the initial start up, an average of 0.2\% with a slightly higher load every two seconds as the constellation is updated.

On the other hand, \textsc{Celestial}'s Machine Manager uses up to 4.5\% of the host's available memory from the start of the simulation, although that number decreases after the demanding initial setup.
Firecracker microVM memory usage increases linearly with the number of booted microVMs, regardless of whether they are suspended or not, as each keeps a \texttt{virtio} memory device that blocks a fixed portion of the host's memory for the VM.
As microVMs are only suspended when their corresponding satellites move out of the bounding box, their memory is not released.
While this has not been an issue in any of our experiments, because microVM memory usage stays below 20\% even on hosts with comparatively little available memory, Firecracker microVMs can also be configured to use ballooning to allow the host to reclaim unused memory from the VMs.

Finally, we also note the cost of our testbed:
For our three hosts and one coordinator, a 10-minute experiment with an additional five minutes for setup and data collection yields a total cost of \$3.30 on Google Cloud Platform.
For comparison, creating 4,409 \texttt{f1-micro} virtual machine instances, with one for each satellite server, costs at least \$539.66 for 15 minutes~\cite{compute_pricing}.

\section{Case Study: Real-Time Ocean Environment Alerts With Remote Sensors}
\label{sec:buoys}

\begin{figure}
    \centering
    \includegraphics[width=\columnwidth]{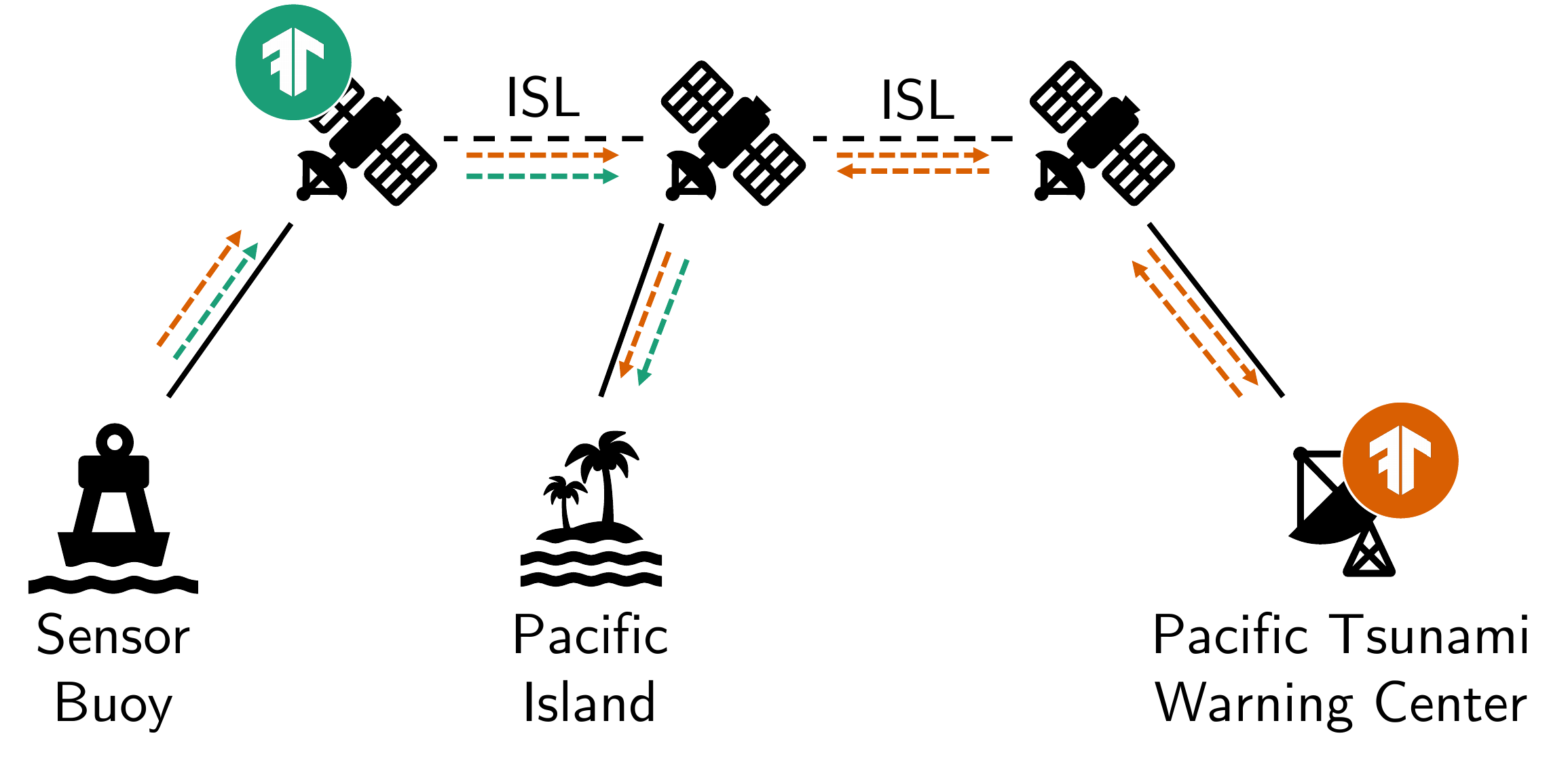}
    \caption{DART stations are remote buoys equipped with environment sensors, such as pressure recorders, that send their measurement data over the Iridium satellite constellation. Sensor data is processed with an LSTM neural network in a central processing location (orange) or on the LEO edge (green). Inferred data is sent to geographically close island ground stations and ships.}
    \Description[DART System Overview]{A DART sensor buoy sends sensor data over the Iridium satellite constellation to a central processing location.}
    \label{fig:dart}
\end{figure}

Using \textsc{Celestial}, we can empirically investigate the potential of the LEO edge for applications that may benefit from running at the network edge.
One class of such applications are real-time monitoring services, e.g., in the context of industrial IoT or environment observation.
The National Oceanic and Atmospheric Administration's (NOAA) \emph{Deep-ocean Assessment and Reporting of Tsunamis} (DART) project uses remote sensing on stationary buoys located in the Pacific to detect early tsunami warning signs~\cite{Gonzalez1998-yo,Meinig2005-qq}.
Given their remote locations, these sensors cannot communicate over terrestrial networks, but instead use the Iridium LEO satellite constellation.
Sensor data is sent to the Pacific Tsunami Warning Center on Ford Island, Hawaii, where it is processed centrally.
This system's architecture is illustrated in \cref{fig:dart}.

\begin{figure}
    \centering
    \includegraphics[width=\columnwidth]{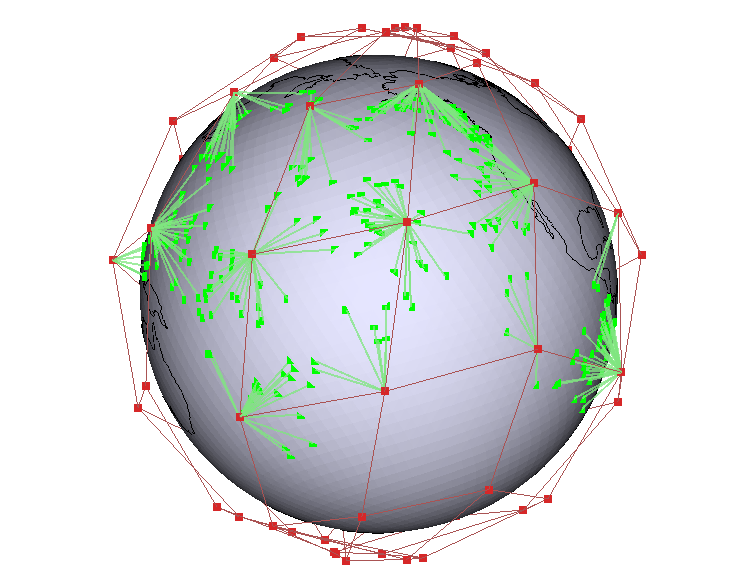}
    \caption{In our experiments, 100 data buoys in the Pacific Ocean send sensor data over the Iridium satellite network.
        Sensor readings are used for inference with a LSTM neural network and results are forwarded to 200 ships and islands.
        Ground stations and their connections are shown in bright green, satellites in red.
        As a result of the 180° arc of ascending nodes in the Iridium constellation, no ISLs exist between satellites of the first and last orbital plane of the constellation, as these satellites move in opposite directions.}
    \Description[Data buoys, islands, and vessels access the Iridium LEO satellite network]{As a result of the 180° arc of ascending nodes in the Iridium constellation, no ISLs exist between satellites of the first and last orbital plane of the constellation, as these satellites move in opposite directions.}
    \label{fig:iridium}
\end{figure}

We use a system inspired by DART to evaluate if and how LEO edge computing can assist real-time ocean environment monitoring for use-cases such as Tsunami warning.
In our experiments, 100 data buoys in the Pacific Ocean transmit sensor readings over the Iridium satellite network.
The Iridium satellite network has a single shell, 66 satellites in 6 planes at a 780km altitude, in a polar orbit (90° inclination) and spaced evenly only around half the globe (180° arc of ascending nodes) so that satellites descending their orbit cover the other half~\cite{De_Weck2004-me}.
As a result of this spacing, the Iridium constellation cannot provide ISLs between the first and last orbital planes, which is reflected in our testbed.
The readings, grouped by sensor location and type, are used to predict weather and environmental events with a long short-term memory (LSTM) neural network~\cite{Hu2019-td}.
Results are distributed to ships and islands in the vicinity of the sensor, using a total of 200 locations.
We show this topology in \cref{fig:iridium}.
We compare two different deployments of the inference service:
First, we deploy it in a central ground station server at the location of the Pacific Tsunami Warning Center on Ford Island, Hawaii.
Second, we deploy the inference service on each of the Iridium satellites, facilitating device-to-device communication.

\begin{figure*}
    \centering
    \begin{subfigure}{0.5\textwidth}
        \centering
        \includegraphics[width=0.9\linewidth]{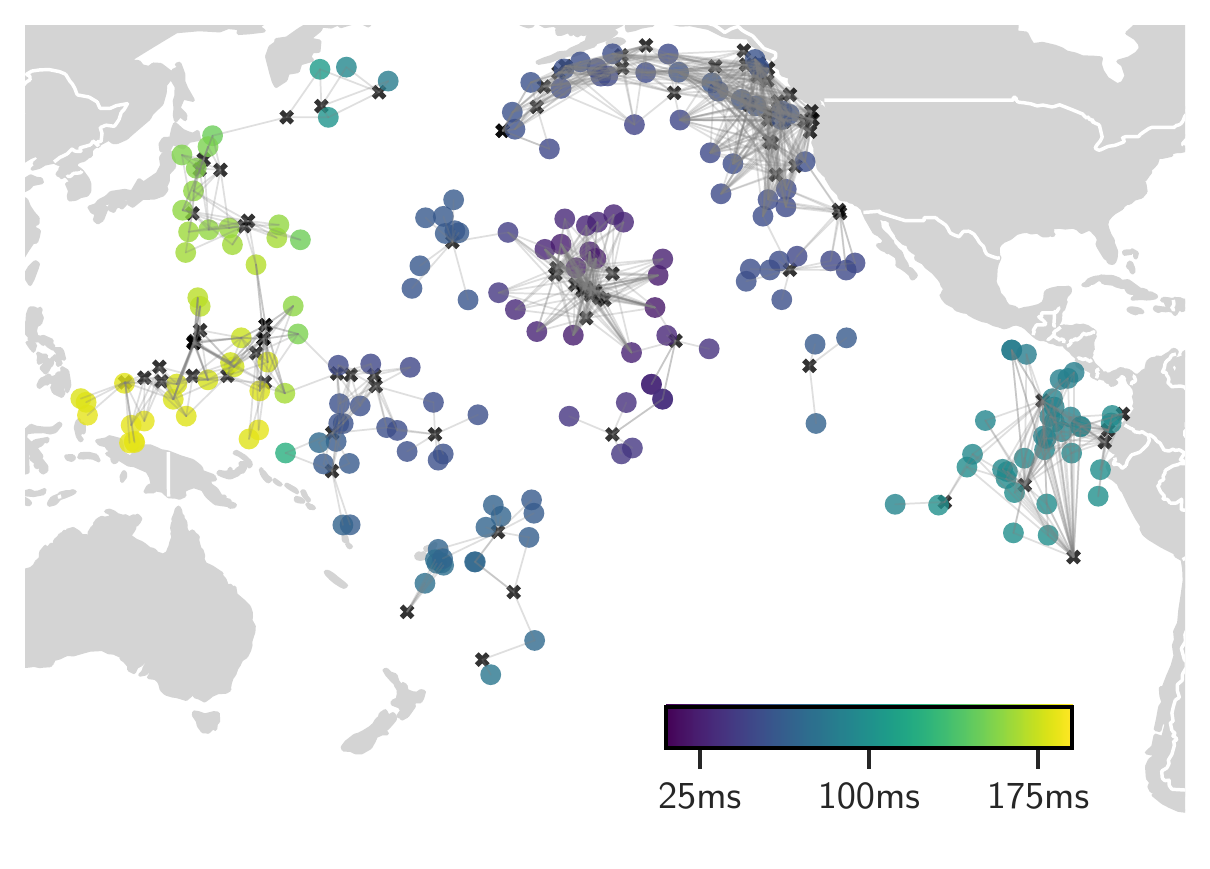}
        \caption{Deployment with Central Processing Location}
        \label{fig:buoys_cloud}
    \end{subfigure}%
    \begin{subfigure}{0.5\textwidth}
        \centering
        \includegraphics[width=0.9\linewidth]{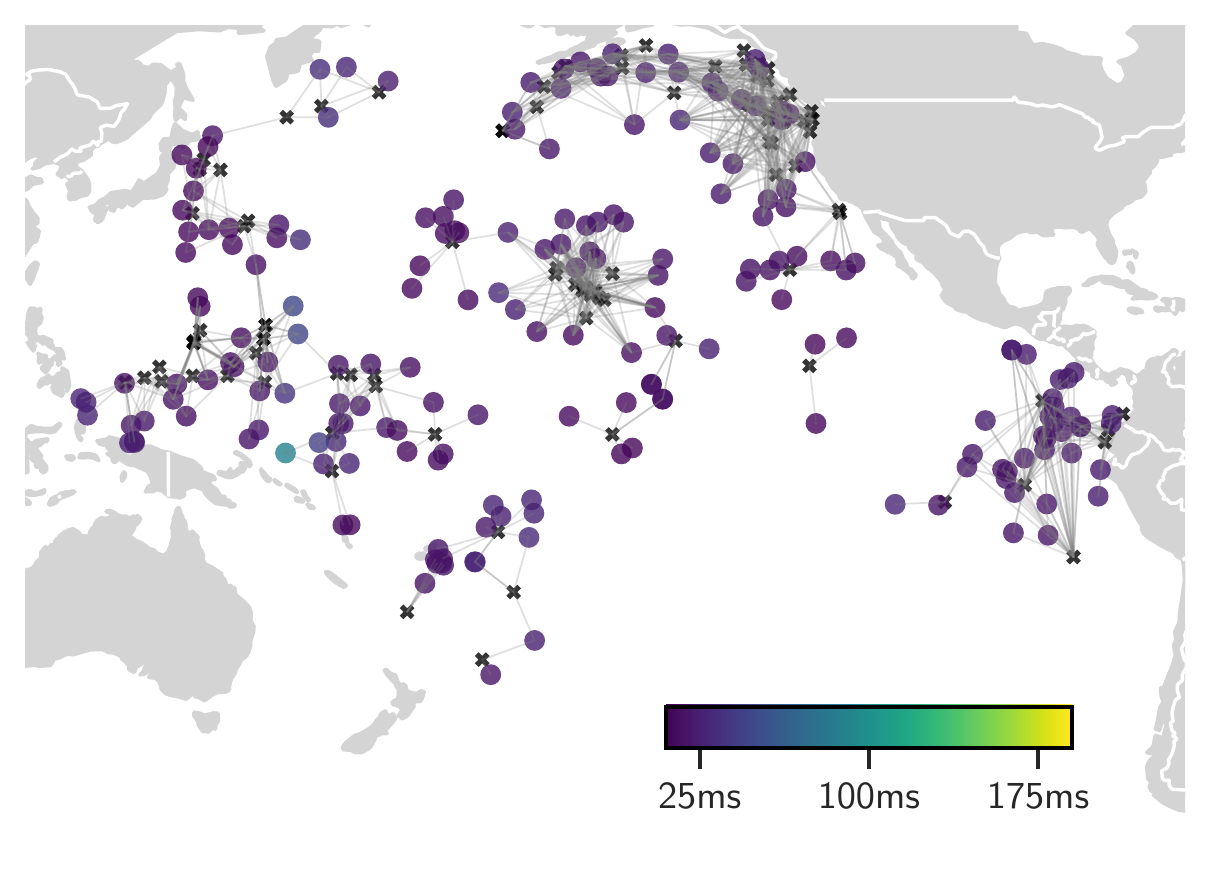}
        \caption{Satellite Server Deployment on Iridium Constellation}
        \label{fig:buoys_sat}
    \end{subfigure}
    \caption{Mean observed end-to-end latency for two different deployments: Colored circles show data sinks, crosses show remote sensor locations, and gray lines indicate data paths. A greater communication distance to the central processing location also increases the communication delay (\cref{fig:buoys_cloud}), while a satellite server deployment results in uniform load distribution (\cref{fig:buoys_sat}). In both deployments, the lack of ISLs between first and last orbital plane of the Iridium constellation increases the communication delay towards the West Pacific region.}
    \Description[Mean end-to-end latency]{Results show increased QoS in the satellite server deployment, except for the area around Hawaii.}
    \label{fig:buoys_results}
\end{figure*}

\subsection{Experiment Setup}

All components of our experiments run on a \textsc{Celestial} testbed.
Sensor data is sent at a one-second interval over UDP.
Again, we use shared PTP clocks to minimize the impact of clock drift.
Both sensors and data sinks are equipped with one CPU core and 1024MB memory.
The inference service uses a TensorFlow stacked LSTM network.
In the satellite deployment, satellite servers each have one CPU core and 1024MB memory, whereas we equip the ground station server with eight CPU cores and 8192MB memory in the datacenter deployment.
Satellite link bandwidth is set at 88Kb/s for sensors and data sinks as recommended by Iridium for remote sensing applications~\cite{iridiumcertus}.
ISLs and links to the central processing ground station are set at 100Mb/s.

All experiments are conducted on a cluster of four GCP \texttt{N2\-highcpu} instances with 32 cores and 32GB memory each, placed in the \emph{europe-west3-c} zone and using an installation of Ubuntu 18.04.
In addition, the Coordinator is hosted on a GCP \texttt{C2} instance with 16 cores and 64GB memory, with an update interval of 5 seconds.
Each experiment is 15 minutes long with an additional five-minute startup phase for the system to stabilize.
We repeat each experiment three times and present results for the median runs.

\subsection{Results}

In \cref{fig:buoys_results}, we show the mean end-to-end latency for the real-time ocean environment alert system.
Overall, the satellite server deployment leads to a better performance compared to the centralized deployment.
As a result of the shorter communication distances, end-to-end latency is reduced from between 22ms and 183ms to between 13ms and 90ms.
Note that processing latency is similar between both deployments, at an average of 2ms.
Further, \cref{fig:buoys_cloud} shows that ground stations that require data from the same sensors observe similar delays, a result of the device-to-device architecture.

The effect of the lack of ISLs between first and last orbital plane of the Iridium constellation can be seen in both deployments:
Over the course of the experiment, locations in the West Pacific region (Asia and Oceania) tend to connect to uplink satellites from the last orbital plane, while those in the Americas connect to the first.
Consequently, any requests between those locations must be routed through satellites near the poles, increasing the communication delay.
In the central processing deployment, this leads to a higher observed latency in this area over the course of the experiment, as all sensor data are routed in this manner.
In higher latitudes, which are nearer to the North Pole, this effect is not as pronounced.
With satellite servers, however, data is only routed across shorter geographical distances.
Increased communication delays can thus only be observed when sensor station and data sink connect to opposite planes of the constellation, which happens less frequently when both are geographically close.

\subsection{Opportunities and Implications for LEO Edge Applications}

The results of our experiments show that LEO edge computing can improve QoS for remote-sensing applications where communication delay is the limiting factor.
Without sending all data to a central processing facility, such as a cloud datacenter, and instead processing it on the communication path, significant latency improvements can be achieved.
For this machine-to-machine communication use-case, on-device processing could not yield such an improvement:
In case of processing on data buoys, sensor data would need to be sent back to buoys of the same group, incurring additional delays.
Processing directly at the data consumers, i.e., island or ship ground stations, requires performing the same computation on several machines.

Please note that our experiment assumed rather small data volumes being sent:
Since communication delay is the main cost factor in end-to-end latency, larger data volumes will increase the latency difference further.
Also, larger data volumes will at some point make it inefficient to transmit all data for centralized processing -- similar to bandwidth limits in terrestrial edge computing~\cite{paper_pallas_fog4privacy}.
Similar to terrestrial edge computing, however, centralized processing may be more efficient for scenarios with low data volumes that can tolerate the higher end-to-end latency.
\section{Discussion \& Future Work}
\label{sec:discussion}

In our evaluation of \textsc{Celestial}, we find that it fulfills our original requirements for a LEO edge testbed.
In this section, we discuss threats to validity as well as limitations of our work and derive avenues for future work on \textsc{Celestial} and the LEO edge.

\subsection{Limits to Scalability}

In our experiments we show that \textsc{Celestial} is horizontally scalable across multiple servers and that we can run an application in a realistic environment of large LEO edge infrastructure.
We use both over-provisioning of our hosts and a bounding box to achieve this in a cost-efficient manner, but there may be use-cases where neither is an option.
When emulation of the entire infrastructure is needed and all satellite and ground station servers run at their full capacity, host infrastructure must be sized appropriately, e.g., requiring 36TB of memory when emulating a full constellation of 4,400 satellites with 8GB memory each.

While \textsc{Celestial} is designed to scale out across as many hosts as necessary, we must assume that such scale comes with caveats.
The network connection between hosts could, for example, become a bottleneck if inter-satellite communication bandwidth exceeds available physical bandwidth.
Some of these effects could be mitigated in \textsc{Celestial} by dynamically migrating satellite server microVMs across hosts to optimize communication and resource provisioning, using a more advanced scheduler such as \emph{FirePlace}~\cite{balaji2021fireplace}.

\subsection{Resource Isolation in microVM Collocation}

Choosing microVMs over alternative technologies allows us to provide an application-agnostic runtime environment while also achieving cost-efficiency through collocation of multiple machines on one physical server.
Nevertheless, that collocation can come at a price if resources are not appropriately available and microVMs start competing for resources.
While satellite servers are independent of each other, with each server placed on an individual satellite, microVMs on one host may affect each other's performance if their processes are scheduled on the same physical host CPU core.
In practice, these effects cannot be easily circumvented without attaching microVM processes to host CPU cores strictly, which limits scalability as it inhibits over-provisioning of hosts.
Yet, this can only become a larger problem once resources required by satellite servers exceed the host's resources, which can be mitigated by scaling \textsc{Celestial} hosts vertically through additional CPU cores or memory, or horizontally by adding further host machines.

\subsection{Impact of microVM Suspension}

\textsc{Celestial}'s bounding box allows for a smaller testbed footprint by suspending microVMs of satellite servers that move outside a specified area.
In our example application, we have seen how this reduces the load on the \textsc{Celestial} hosts without impacting the application.
There may be some cases where microVM suspension has unwanted side effects on the user's application.
If the satellite server software is expected to change its state based on its location, e.g., a CDN service that needs to proactively replicate files before it is in reach of a ground station, it will not be able to do so while the machine is suspended and will potentially have to catch up to missed updates once it is activated again.
To avert this, a user may increase the size of their bounding box, possibly to cover the entire earth so that no microVM is ever suspended.
Note that this requires more host resources and hence increases cost.

\subsection{Assumptions on Hardware Architecture}

To work on commodity hardware, \textsc{Celestial} assumes satellite servers to have the same hardware architecture as common terrestrial servers, i.e., the \emph{x86-64} architecture supported by Firecracker.
In reality, we cannot currently know the hardware which will be used on the LEO edge.
Specific hardware features such as special instructions or real-time guarantees can thus not be emulated on \textsc{Celestial} at the moment and would require full hardware virtualization or at least emulation of a subset of instructions.
Still, even if future LEO edge operators were to go the unlikely path of developing a unique compute infrastructure for satellite servers, \textsc{Celestial} could still provide a realistic environment to evaluate algorithms and programming models.

\subsection{Emulating External Factors}

Although \textsc{Celestial} can emulate satellite server degradation, e.g., caused by radiation, there are many other external factors that can impact satellite constellations.
To give one example, satellites may not always follow completely deterministic orbits.
Starlink's satellites are equipped with ion thrusters to adjust their orbits, which can be necessary to dodge space debris or other satellites~\cite{8594798,ALFANO2021241}.
This has effects on the network as well, as physical distances change or ISLs can become unavailable during such maneuvers.

Adverse weather conditions can have an effect on the radio link between a ground station and satellites, with radio dishes overheating~\cite{Brodkin2021-dh} and rain causing refraction of radio waves~\cite{Safaai-Jazi1995-ei}.
Ground station equipment may also be mobile, e.g., if installed on a plane or car, which must be taken into account when selecting uplink satellites.
Such factors may impact LEO edge applications, and it can be helpful to test their effects ahead of application deployment.

The separation of Constellation Calculation and Machine Managers in \textsc{Celestial} will allow researchers to incorporate new models quickly, as only the calculation component must be extended.
The resulting networking and machine parameters can then be sent to the host machines without modification.
Additionally, \texttt{tc} and \texttt{tc-netem} offer advanced network emulation features that can be used in \textsc{Celestial} in the future with only small changes to its codebase, such as packet loss or duplication, delay distributions, packet corruption, or packet reordering.
Whether these emulated characteristics will be useful for LEO edge testbeds depends on future research on LEO satellite network measurements.

While no efficiently emulated testbed can ever be fully accurate with respect to the environment it emulates, we believe that the current feature set of \textsc{Celestial} accurately reflects what is known about the LEO edge today.
As future research on LEO satellite networks will likely yield additional insights on the effects of external factors, \textsc{Celestial} can easily be adapted or extended in all possible directions thanks to its small and simple codebase.

\subsection{Lack of Validation Data}

With \textsc{Celestial}, one of our goals is to emulate the LEO edge accurately, yet the only method to verify this is to compare our measurements with simulation results.
What the research community is still missing are realistic data traces for large LEO satellite constellations, as network operators guard their data to keep a competitive edge.
The recent \emph{SatNetLab} proposal for a research platform for global satellite-based networks~\cite{singla2021satnetlab} may change this in the future, and we hope to adapt \textsc{Celestial}'s network calculation as new data becomes available.

\subsection{Future Work on LEO Edge Systems}
\label{subsec:future_leo_work}

With its scale and dynamic topology, the LEO edge poses significant challenges for applications.
To simplify the deployment of services, systems researchers should address these challenges with application platforms that abstract from the underlying infrastructure.
The number of satellite servers will require new coordination approaches to guarantee service consistency.
Further, the servers' limited resources must be allocated efficiently, while they are possibly even shared by multiple tenants~\cite{Pfandzelter2021-dp}.

The high mobility of LEO satellites introduces the additional challenge of state management:
Clients will frequently need to connect to a new satellite edge server, and any server-side state must be migrated accordingly.
Bhattacherjee et al.~\cite{Bhattacherjee2020-kr} have proposed the concept of ``virtual stationarity'', where such state is migrated between satellites based on their locations relative to Earth, so that data appears to be in the same location from a client perspective.
Further, such state management requires routing the clients' requests to the correct satellite server, which is made more difficult by the dynamic network topology.

\textsc{Celestial} itself does not include any strategies for state management, request routing, or service management, as it is intended as a testbed on which future systems implementing such strategies can be evaluated.

\subsection{Feasibility of the LEO Edge}

Finally, \textsc{Celestial}'s utility depends heavily on the future of the LEO edge.
While considerable resources are committed to the development of large LEO satellite networks to provide global Internet coverage, LEO edge computing has so far only received limited attention from research and industry.
If current trends in this field do not come to fruition and satellite network operators do not see sufficient financial incentives in operating LEO compute infrastructure, \textsc{Celestial} can only be used for theoretical evaluation.

Nevertheless, we also see this as an opportunity to discover those incentives:
\textsc{Celestial} makes it possible to easily evaluate possible LEO edge applications and to bootstrap the development of LEO edge infrastructure.

\section{Conclusion}
\label{sec:conclusion}

In this paper, we have motivated the need for a testbed for the LEO edge that enables systems researchers, application developers, and platform designers to test and evaluate their LEO edge computing software on Earth.
With \textsc{Celestial}, we have answered how such a testbed can be built for accuracy, in an application-agnostic manner, and with cost-efficiency and scalability in mind.
In support of those claims, we have used \textsc{Celestial} to deploy a LEO edge application from the existing body of research.
Additionally, we have shown in a case study how we use \textsc{Celestial} to empirically evaluate the potential of the LEO edge for remote sensing applications.

Finally, we have laid out interesting avenues for future work:
With \textsc{Celestial}, researchers will now be able to build and test platforms and systems for the LEO edge that address the challenges of state management, resource allocation, or request routing.
To further improve the accuracy of emulated testbeds, experiences and data of real LEO edge deployments are needed.
We hope that \textsc{Celestial}, which we have published as open-source, can be of great use in this further research on and development of LEO edge computing.

\balance

\begin{acks}
    We thank our anonymous reviewers, our anonymous shepherd, and our colleague Dr.~Jonathan Hasenburg for their valuable feedback on this work.
    Funded by the \grantsponsor{DFG}{Deutsche Forschungsgemeinschaft (DFG, German Research Foundation)}{https://www.dfg.de/en/} -- \grantnum{DFG}{415899119}.
    This work is supported by the Google Cloud Research Credits program (GCP202443755) and by the AWS Cloud Credit for Research program.
\end{acks}

\bibliographystyle{ACM-Reference-Format}
\bibliography{bibliography}

\end{document}